\documentclass[11pt]{article}
\usepackage[margin=1in]{geometry}
\usepackage{graphicx}

\def\beq{\begin{equation}}
\def\eeq#1{\label{#1}\end{equation}}
\def\eeqn{\end{equation}}

\def\beqa{\begin{eqnarray}}
\def\eeqa#1{\label{#1}\end{eqnarray}}
\def\eeqan{\end{eqnarray}}

\let\bar=\overbar

\def\Dslash{\not{\hbox{\kern-4pt $D$}}}
\def\dslash{\not{\hbox{\kern-2pt $\del$}}}

\def\msb{{\bar{\ssstyle M \kern -1pt S}}}

\def\Title#1{\begin{center} {\Large {\bf #1} } \end{center}}
\def\Author#1#2{\begin{center} {\normalsize {\sc #1}}\\{\small #2} \end{center}}
\def\Institution#1{\begin{center} {\normalsize {\it #1} } \end{center}}
\def\Abstract#1{\noindent {\normalsize {\bf Abstract:} {\normalfont #1}}}
\def\Conference{\vspace{4mm}\begin{raggedright} {\normalsize {\it Talk presented at the 2019 Meeting of the Division of Particles and Fields of the American Physical Society (DPF2019), July 29--August 2, 2019, Northeastern University, Boston, C1907293.} } \end{raggedright}\vspace{4mm}}

\usepackage{amsmath}
\usepackage{xspace}

\newcommand{\ns}{\ensuremath{\,\text{ns}}\xspace}
\newcommand{\GeV}{\ensuremath{\,\text{GeV}}\xspace}
\newcommand{\cms}{\ensuremath{\,\text{cm}^{-2}\text{s}^{-1}}\xspace}
\newcommand{\pt}{\ensuremath{p_{\text{T}}}\xspace}
\newcommand{\mm}{\ensuremath{\,\text{mm}}\xspace}
\newcommand{\um}{\ensuremath{\,\mu\text{m}}\xspace}

\begin{document}

%
%

\Title{Level 1 Track Finder at CMS}

\Author{Andrew Hart}{for the CMS Tracker Group}

\Institution{Department of Physics and Astronomy\\ Rutgers, The State University of New Jersey, Piscataway, NJ, USA}

\Abstract{The High-Luminosity LHC is expected to deliver proton-proton collisions every 25\ns with an estimated 140--200 pileup interactions per bunch crossing. Ultrafast track finding is vital for handling Level 1 trigger rates in such conditions. An FPGA-based track trigger system, capable of finding tracks with momenta above 2\GeV, is presented.}

\Conference

%
%

\section{Introduction}

The High-Luminosity LHC is expected to achieve luminosities up to $5 \times
10^{34}\cms$, or up to $7.5 \times 10^{34}\cms$ in the ultimate performance
scenario. This amount of data offers great opportunities in terms of potential
physics results. However, the pileup associated with this luminosity is at the
level of 140--200 interactions per bunch crossing, presenting serious
challenges at all stages of data collection and analysis.

\begin{figure}[htb]
\centering
\includegraphics[width=0.49\textwidth]{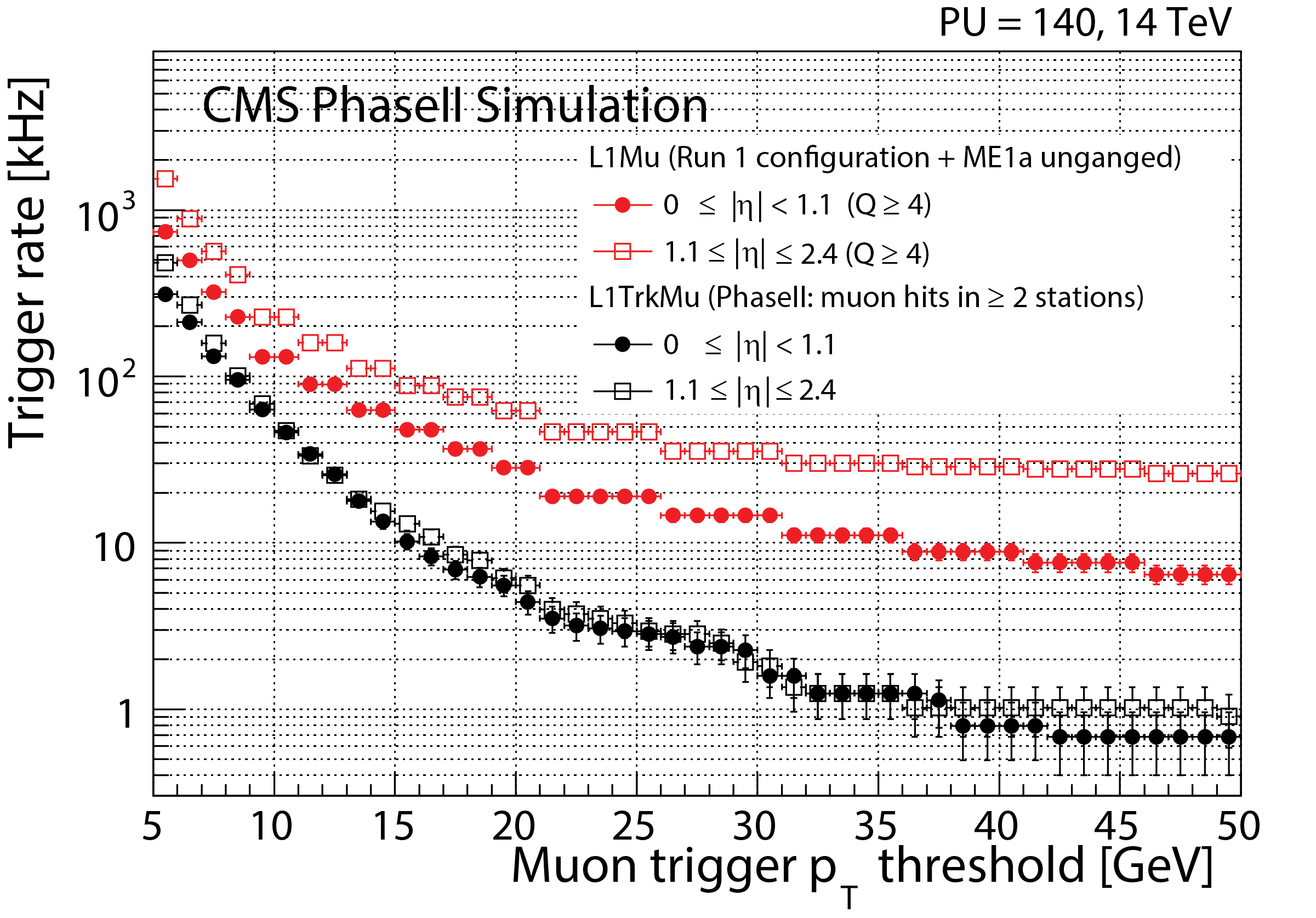}
\includegraphics[width=0.49\textwidth]{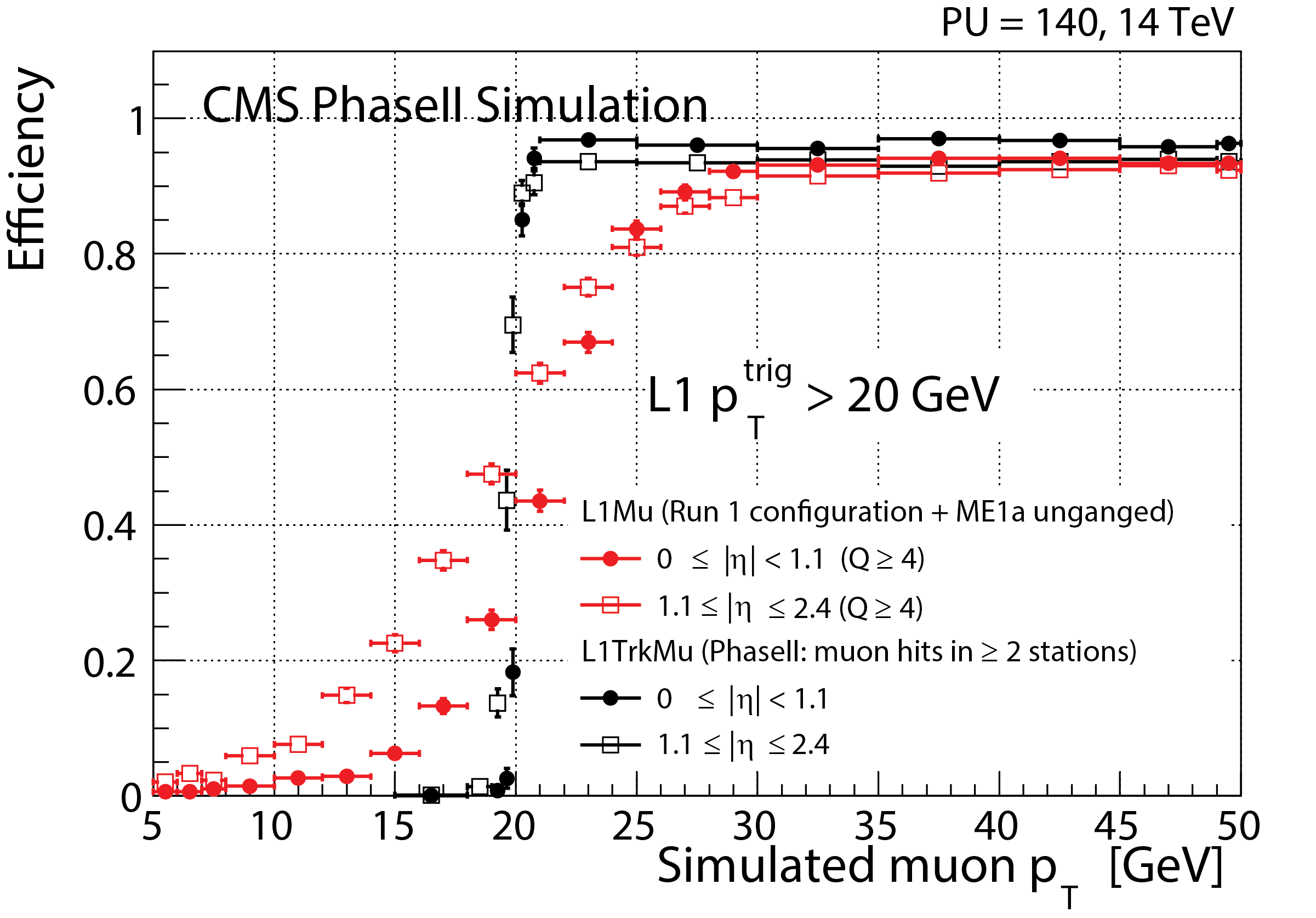}
\caption{Trigger rates versus muon trigger \pt threshold (left) and trigger
  efficiencies versus simulated muon \pt (right) for a single-muon trigger
  without (red points) and with (black points) L1
  tracks~\cite{CMSCollaboration:2015zni}.}
\label{fig:muons}
\end{figure}

Providing tracks to the Level 1 (L1) trigger~\cite{Khachatryan:2016bia} is a
key part of the strategy that CMS~\cite{Chatrchyan:2008aa} will employ to cope
with the high amounts of pileup~\cite{CMSCollaboration:2015zni,Klein:2017nke}.
Tracks at L1 will help mitigate the effects of pileup in several ways, e.g.,
increasing the purity of L1 muons by requiring an associated L1 track, thus
reducing background rates (see the left plot of Figure~\ref{fig:muons}). They
will also help improve the measurement of muons and other objects that have
tracks (see the right plot of Figure~\ref{fig:muons}). Finally, L1 tracks open
up possibilities for new kinds of triggers that are currently impossible, e.g.,
those based on displaced or disappearing tracks~\cite{CMS:2018qgk}.

\begin{figure}[htb]
\centering
\includegraphics[width=0.52\textwidth]{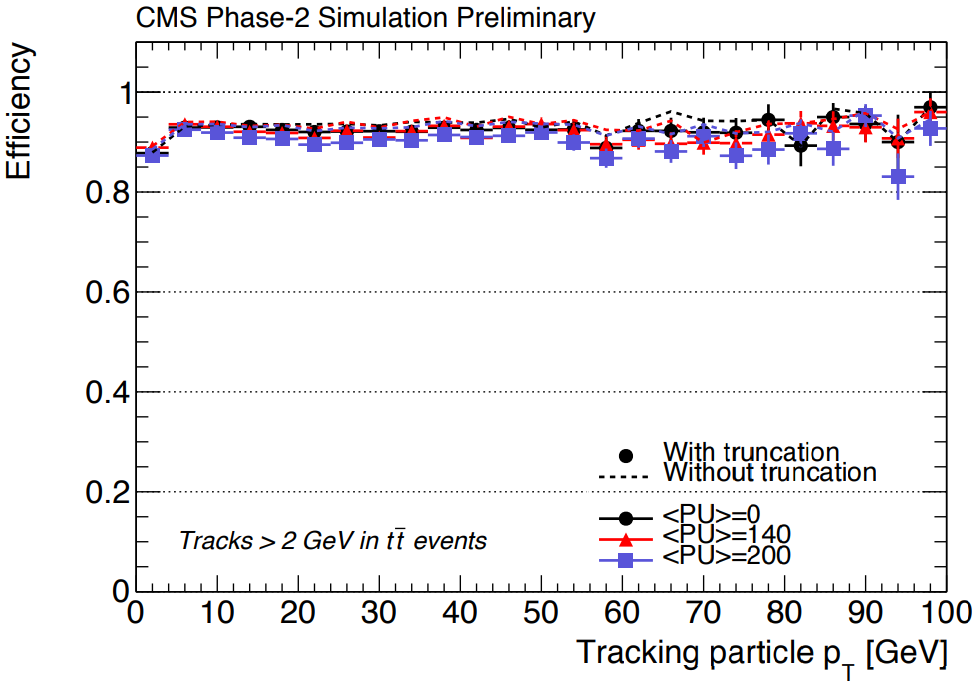}
\includegraphics[width=0.46\textwidth]{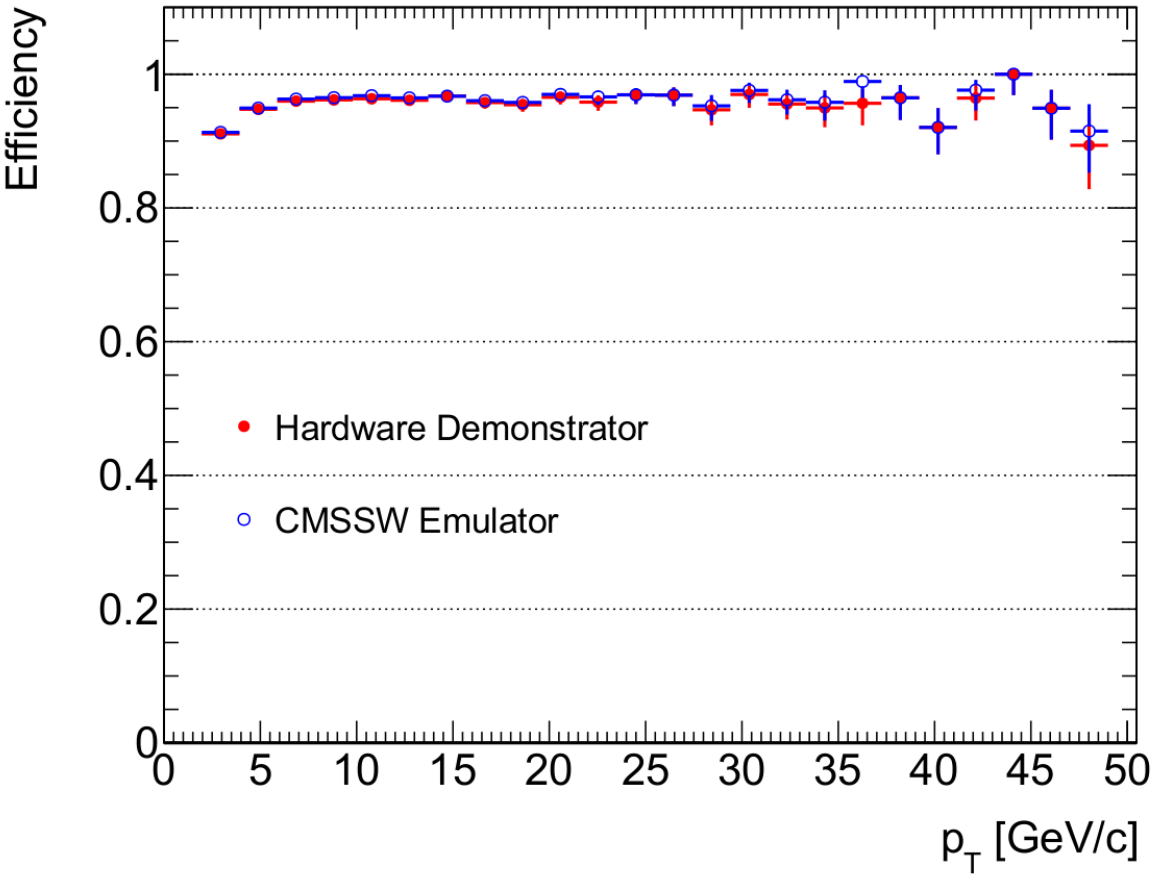}
\caption{L1 tracking efficiencies versus simulated particle \pt for the
  tracklet algorithm (left) and the TMTT algorithm (right).}
\label{fig:efficiencies}
\end{figure}

\begin{figure}[htb]
\centering
\includegraphics[width=0.52\textwidth]{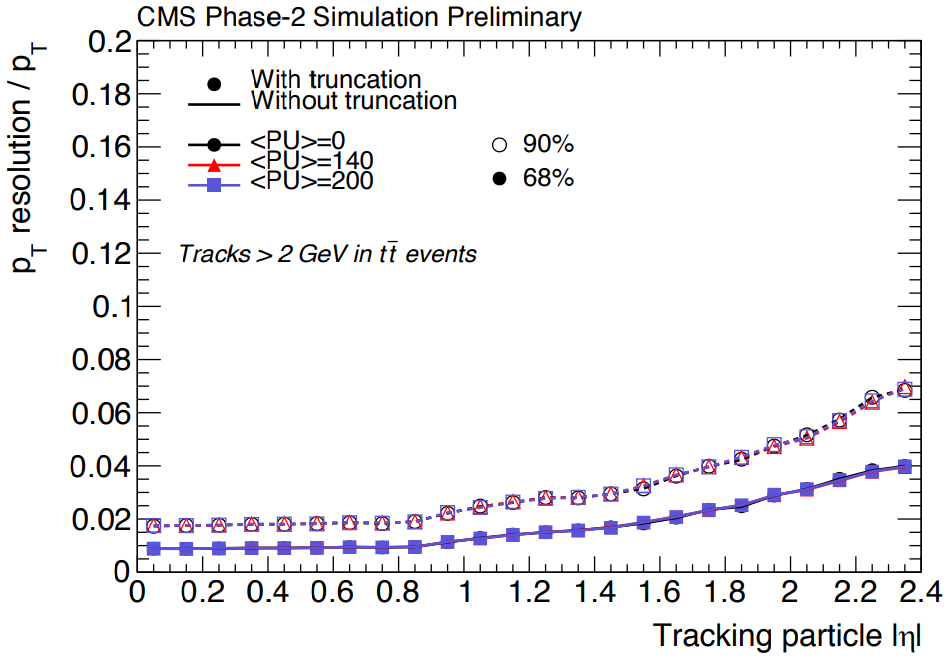}
\includegraphics[width=0.46\textwidth]{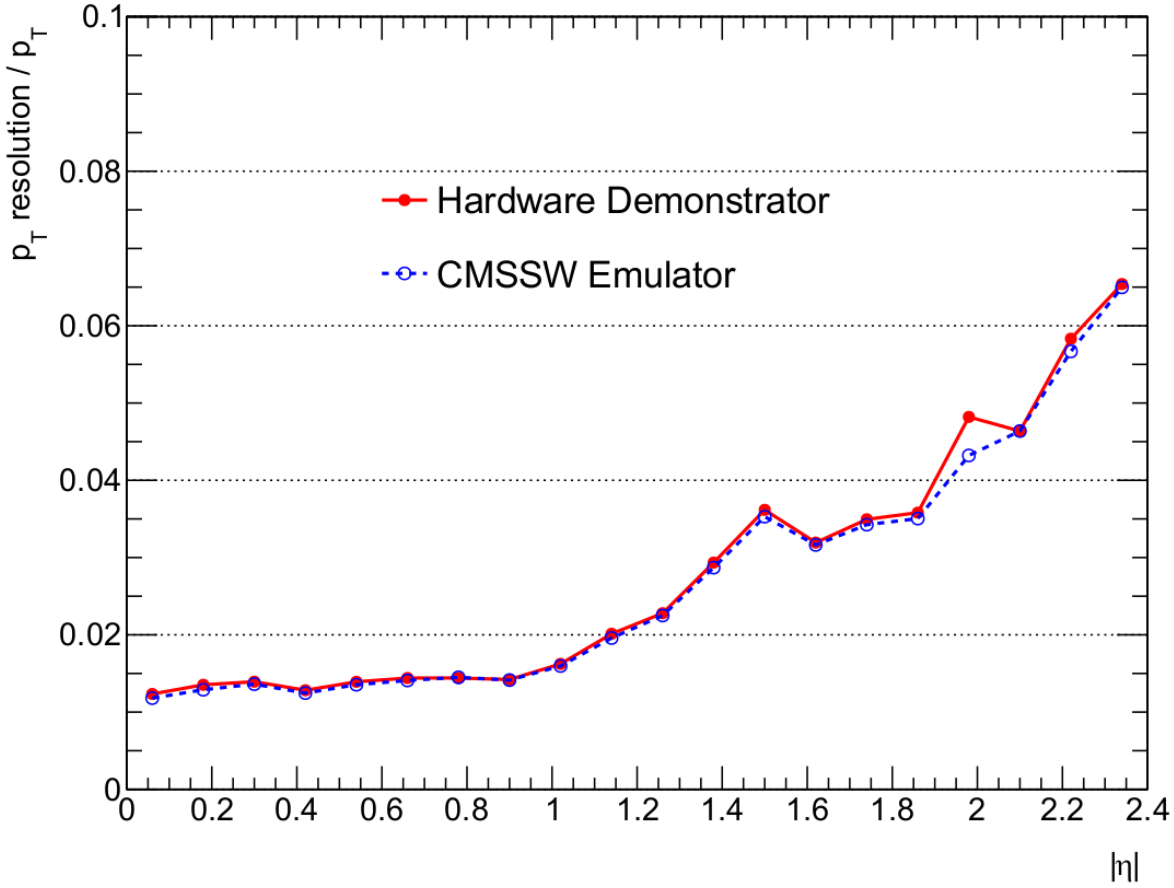}
\caption{Relative transverse momentum resolutions versus simulated particle
  $|\eta|$ for L1 tracks produced by the tracklet algorithm (left) and the TMTT
  algorithm (right).}
\label{fig:resolutions}
\end{figure}

Two all-FPGA track trigger algorithms have been developed by CMS, which differ
in their approaches to both pattern recognition and track fitting.  The
\emph{tracklet} algorithm employs a straightforward road search for pattern
recognition, followed by a simple, linearized $\chi^2$ fit. On the other hand,
the \emph{time-multiplexed track finder} (TMTT) algorithm uses a Hough
transform to find tracks, and a Kalman filter to fit the tracks that are found.
Both approaches have achieved similar track-finding efficiencies and track
parameter resolutions in emulation, as can be seen in
Figures~\ref{fig:efficiencies} and \ref{fig:resolutions}. Furthermore,
technical demonstrations in 2016 proved the feasibility of both approaches in
hardware.

The current focus now is on a hybrid algorithm that combines the most
sophisticated parts of the two approaches: using a road search for pattern
recognition and a Kalman filter for track fitting. This algorithm will be
outlined in this talk.

\section{Track trigger algorithm}

\begin{figure}[htb]
\centering
\includegraphics[width=0.6\textwidth]{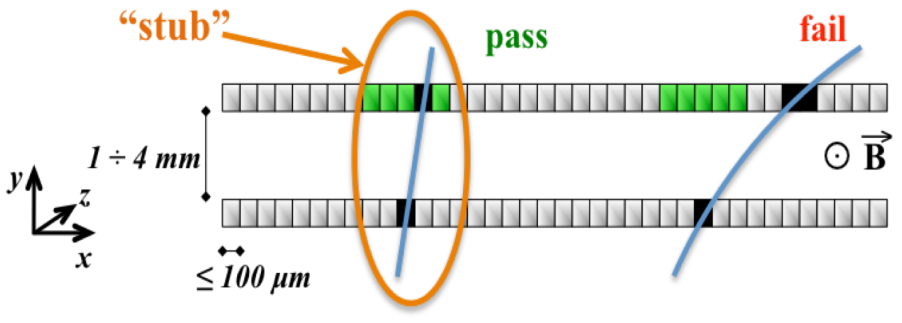}
\caption{Illustration of the \pt discrimination capabilities of the two-sided
  \pt modules. The stub on the left is consistent with the chosen \pt threshold
  and is read out by the front-end, while the stub on the right is not.}
\label{fig:ptModules}
\end{figure}

Track finding begins with track stubs that are formed by two types of \pt
modules, each of which contains two layers of active material with a small gap
between. In the pixel-strip modules, one of the layers is composed of $1.5\mm
\times 100\um$ pixels while the other layer is a strip sensor with a $100\um$
strip pitch. These modules will be used closer to the interaction point, where
the higher occupancy demands finer segmentation. Farther from the interaction
point, strip-strip modules will be used, where both layers of active material
are strip sensors with a pitch of $90\um$. The two-sided nature of the modules
allows for front-end \pt discrimination. As illustrated in
Figure~\ref{fig:ptModules}, stubs with too low \pt are rejected, which results
in a data reduction factor of 10--100.

\begin{figure}[htb]
\centering
\includegraphics[width=0.75\textwidth]{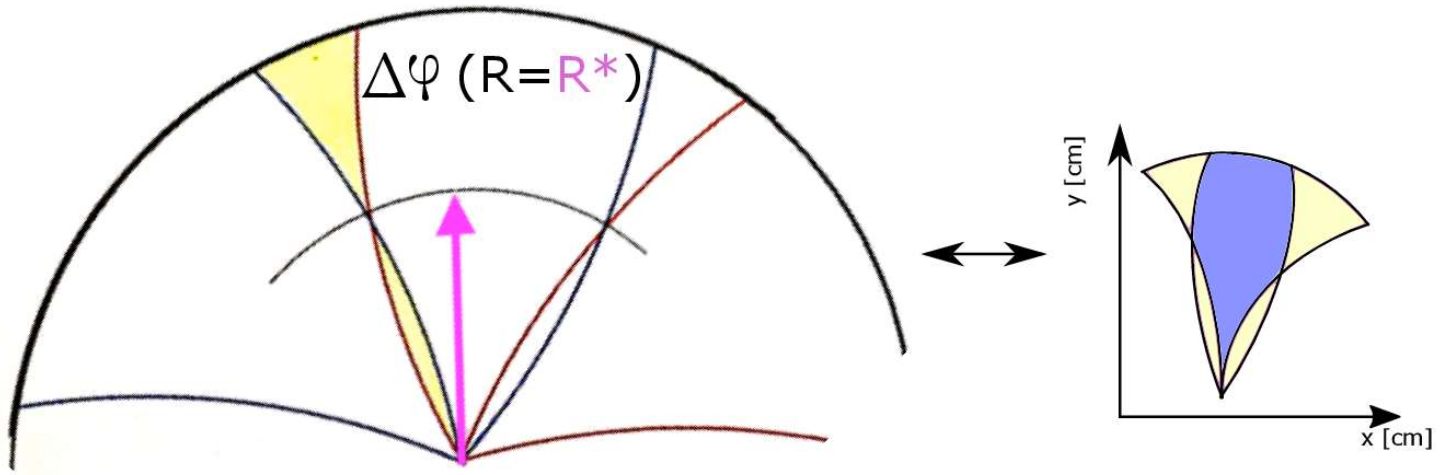}
\caption{Hourglass sector shape used to divide the tracker for the purpose of
  parallelizing the track-finding algorithm. The yellow region is shared by
  neighboring sectors, and stubs in this region must be duplicated. However,
  the curved edges of the sector are such that tracks with \pt above a certain
  threshold can appear in only one sector.}
\label{fig:hourglassSectors}
\end{figure}

The track-finding algorithm that proceeds after stub formation is parallelized,
both in time and space. It is time-multiplexed with a factor of 18 in the
current design, and the tracker is divided into nine ``hourglass'' sectors,
with an independent instance of the algorithm processing the stubs from each
sector. The hourglass shape, shown in Figure~\ref{fig:hourglassSectors},
prevents tracks with a \pt above a certain threshold from entering more than
one sector, thus eliminating the need for cross-sector communication of tracks.
The critical radius ($R^*$ in the figure), is a parameter that is tuned to
minimize the overlap regions between sectors in which stub data must be
duplicated.

\begin{figure}[htb]
\centering
\includegraphics[width=0.25\textwidth]{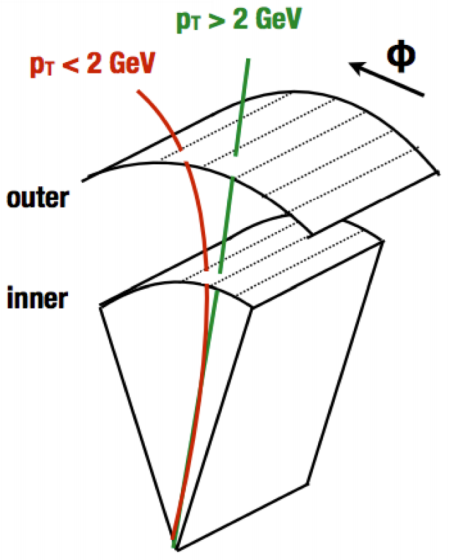}
\caption{Illustration of the \pt discrimination applied when forming stub
  pairs. The green track produces stubs in VMs that are considered for pairing,
  while the red track does not.}
\label{fig:virtualModules}
\end{figure}

Once stubs are formed, track finding begins in each sector by finding pairs of
stubs in adjacent tracker layers that are consistent with a track. These stub
pairs act as seeds from which full tracks are grown. To reduce the volume of
data that has to be processed in subsequent steps, only stub pairs consistent
with $\pt > 2\GeV$ are kept. This is achieved by coarsely segmenting the
tracker layers into virtual modules (VM), 16 or 32 per layer per sector, as
illustrated in Figure~\ref{fig:virtualModules}. Then, only VM pairs consistent
with the \pt threshold of 2\GeV are connected in the firmware, with all other
combinations being ignored. Similarly, the tracker is segmented into eight bins
in the longitudinal direction, and only stubs in combinations of bins that are
consistent with a track originating from near the nominal interaction point are
considered for pairing.

\begin{figure}[htb]
\centering
\includegraphics[width=0.30\textwidth]{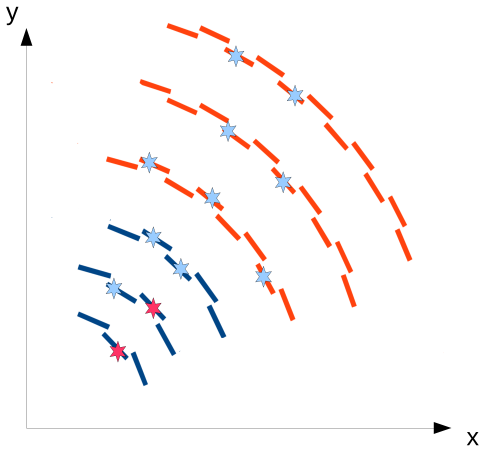}
\includegraphics[width=0.30\textwidth]{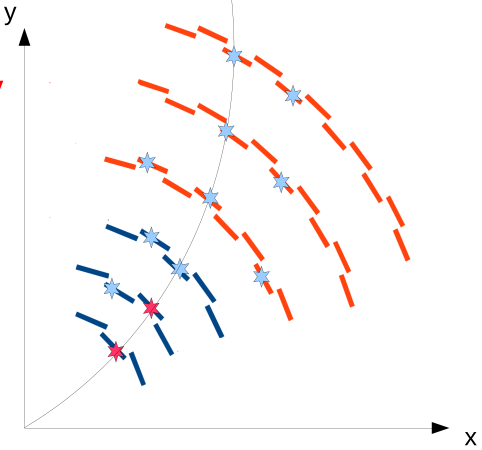}
\includegraphics[width=0.30\textwidth]{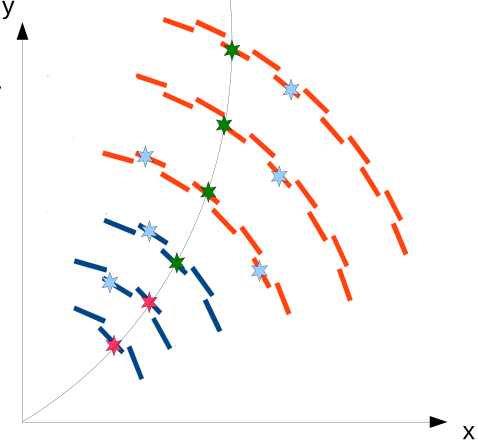}
\caption{Illustrations of the processes of seeding, calculating projections,
  and matching additional stubs. The red stars in the two innermost layers of
  the leftmost illustration indicate a stub pair that will be used to seed a
  track. Projections to the other four layers are calculated in the middle
  illustration. Finally, the green stars in the rightmost illustration indicate
  additional stubs that are matched to the track based on their residuals with
  respect to the projections.}
\label{fig:stubMatching}
\end{figure}

After the track seeds are found, the helix parameters and projections to other
tracker layers are calculated for these ``tracklets,'' assuming they originate
from the beamline. The projections are used to calculate residuals and match
stubs in additional layers, which yields full tracks that can then be fit. The
processes of seeding, calculating projections, and matching additional stubs
are illustrated in Figure~\ref{fig:stubMatching}.

\begin{figure}[htb]
\centering
\includegraphics[width=0.30\textwidth]{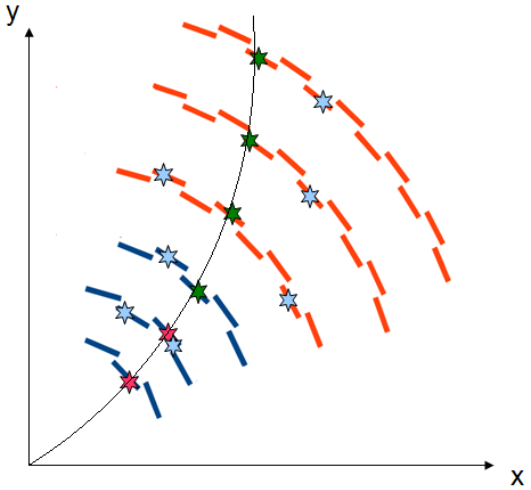}
\includegraphics[width=0.30\textwidth]{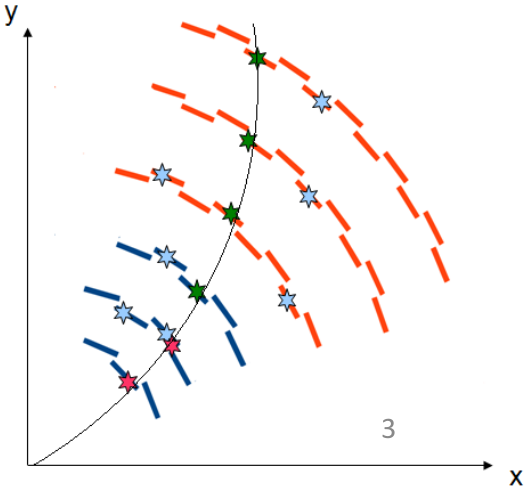}
\caption{Two distinct tracks resulting from the same charged particle. The two
  tracks are identical except for the stub in the second layer, where there is
  ambiguity.}
\label{fig:duplicateTrack}
\end{figure}

However, the pattern recognition naturally produces several duplicate tracks
for a given charged particle. Most come from redundancies in the seeds that are
used to maximize track-finding efficiency; i.e., a given charged particle will
typically be seeded multiple times in different pairs of adjacent tracker
layers, resulting in multiple tracks for the same particle. A few also come
from nearby stubs in a given layer yielding very similar tracks, a scenario
illustrated in Figure~\ref{fig:duplicateTrack}. Whatever the origin, these
duplicate tracks have to be removed before track fitting to reduce the
processing required for that step, and the current strategy is to merge any
tracks that share at least four stubs, although this is an active area of
development.

The final fit of the tracks is done with a Kalman filter. The filter starts
with the coarse helix parameters that were calculated previously for the
tracklet seed. Then stubs are added one by one, as the helix parameters are
updated with greater and greater precision. Currently, there is a beamline
constraint and only four parameters are fit. However, the possibility of
removing this constraint and also fitting for the transverse impact parameter
is being investigated.

\section{Firmware status}

For firmware development, we have chosen to employ Vivado High-Level Synthesis
(HLS) from Xilinx. This allows FPGA designs to be written in C++ instead of a
traditional HDL such as Verilog or VHDL. This enables more rapid development,
especially by individuals who may not have experience writing firmware. Also,
the result should be easier to maintain, and new ideas can be prototyped more
easily.

The current design is divided into nine processing modules, with multiple
instances of each module being employed in the design, and with memories used
to communicate the results between steps. Nearly all of these have at least one
instance written and tested to be functionally correct, and additional
instances will be generated using C++ template programming. Of those that have
been written, nearly all have achieved the desired pipelining, and about half
(four of the nine modules) have been fully verified with C/RTL cosimulation.
This means that, for these modules, the RTL generated by Vivado HLS produces
results that agree exactly with the C++ source code. The goal is to have a full
chain of modules ready for integration tests at CERN in the autumn of 2019.

\section{Conclusion}

A common Level 1 tracking algorithm for the CMS upgrade for the High-Luminosity
LHC is under development. It is a hybrid algorithm based on the most
sophisticated aspects of two proven all-FPGA approaches: tracklet and
time-multiplexed track finder. Development of the firmware with Vivado
High-Level Synthesis is well underway with about half of the processing steps
having fully verified modules written, with a full chain expected to be ready
for integration tests at CERN in the autumn of 2019.

\end{document}